\documentclass[journal=jacsat,preprint]{achemso}

\usepackage[utf8]{inputenc}
\usepackage{amssymb}
\usepackage{graphicx}
\usepackage{caption}
\usepackage{subcaption}
\usepackage{mhchem}
\usepackage{array}
\usepackage{tabularx}
\usepackage{float}
\usepackage{xcolor}
\usepackage{pdfpages}
\usepackage{soul}
\usepackage{fancyhdr}
\usepackage{hyperref}

\sethlcolor{cyan}

\setkeys{acs}{
  articletitle = true,
  etalmode     = truncate,
  maxauthors   = 10,
  biblabel     = fullstop
}

\title{Direct Simulation of \ce{LiNi_{0.8}Mn_{0.1}Co_{0.1}O_{2}} Transport Properties Using an Efficient and Accurate Machine Learning Potential}

\author{Jian He}
\affiliation{Materials Chemistry and Catalysis, Debye Institute for Nanomaterials Science, Utrecht University, Universiteitsweg 99, 3584 CG Utrecht, The Netherlands}
\author{Constantijn H. J. A. van de Wetering}
\affiliation{Materials Chemistry and Catalysis, Debye Institute for Nanomaterials Science, Utrecht University, Universiteitsweg 99, 3584 CG Utrecht, The Netherlands}
\author{Rolande W. Nolsen}
\affiliation{Materials Chemistry and Catalysis, Debye Institute for Nanomaterials Science, Utrecht University, Universiteitsweg 99, 3584 CG Utrecht, The Netherlands}
\author{Nongnuch Artrith}
\affiliation{Materials Chemistry and Catalysis, Debye Institute for Nanomaterials Science, Utrecht University, Universiteitsweg 99, 3584 CG Utrecht, The Netherlands}
\email{n.artrith@uu.nl}

\begin{document}

\maketitle

\section*{Abstract}
The rate capability of layered lithium nickel manganese cobalt oxide (NMC) cathode materials plays a decisive role in high-power applications such as fast charging, necessitating a detailed understanding of lithium-ion diffusion. However, the mechanisms governing lithium-ion transport in NMC remain insufficiently understood, both experimentally and computationally. In this study, we employ an advanced and efficient machine learning potential (MLP) to simulate lithium self-diffusion in \ce{LiNi_{0.8}Mn_{0.1}Co_{0.1}O_{2}} (NMC811), enabling direct large-scale molecular dynamics (MD) simulations. The workflow integrates a fine-tuned MACE (Message Passing Atomic Cluster Expansion) foundation model as a structural generator and leverages an active learning strategy applied to a near-ground-state dataset. This approach enables the construction of a reliable MLP for NMC811 in a data-efficient manner using a limited number of density functional theory (DFT) reference calculations. Based on this potential, we performed MD simulations to predict lithium diffusion coefficients. The MLP-based simulations preserve the accuracy of DFT while overcoming its time and length scale limitations, thereby allowing direct simulation of lithium self-diffusion in NMC811.

\section*{Introduction}
To meet the growing demand for high-energy-density lithium-ion batteries (LIBs), current layered cathode materials face significant challenges, including capacity degradation and instability at the cathode–electrolyte interface, particularly under high-rate operating conditions.

Among these, \ce{LiNi_{0.8}Mn_{0.1}Co_{0.1}O_{2}} (NMC811) has emerged as a promising cathode material for next-generation LIBs due to its high energy density and cost effectiveness.\cite{nohComparisonStructuralElectrochemical2013,linHierarchicalNickelValence2021,liHighnickelLayeredOxide2020} However, its intrinsically disordered structure poses substantial challenges for experimental characterization of microscopic properties. Recent studies have confirmed that hindered lithium diffusion is a key factor contributing to capacity fade under high-rate conditions.\cite{kasnatscheewLearningElectrochemicalData2017,mcclelland_direct_2023,cmcnulty_understanding_2023}

A fundamental understanding of lithium diffusion in NMC811 is therefore critical for improving rate performance. Yet, experimental measurements of lithium diffusivity, particularly as a function of state of charge (SOC), suffer from large uncertainties. For instance, galvanostatic intermittent titration technique (GITT), electrochemical impedance spectroscopy (EIS), and potentiostatic intermittent titration technique (PITT) often yield lithium self-diffusion coefficients that differ by several orders of magnitude.\cite{nohComparisonStructuralElectrochemical2013,tubtimkunaDiffusionZirconiumIV2022,mcclelland_direct_2023,markerEvolutionStructureLithium2019e,chienRapidDeterminationSolidstate2023}

Computational methods play an essential role in probing lithium transport. However, most DFT-based studies are restricted to evaluating energy barriers of specific diffusion mechanisms in small supercells due to prohibitive computational cost. For example, in layered \ce{LiCoO2} (LCO), the oxygen dumbbell hop (ODH) and tetrahedral site hop (TSH) mechanisms have been extensively studied,\cite{vandervenFirstprinciplesTheoryIonic2001,venLithiumDiffusionLayered2000,vandervenLithiumDiffusionMechanisms2001} and similar mechanisms have been assumed for NMC materials.\cite{hoangDefectPhysicsChemistry2016,jaberiStudyLithiumTransport2024,dixitThermodynamicKineticStudies2016} Nevertheless, the disordered nature of NMC811 suggests that diffusion pathways may vary significantly, motivating the use of direct MD simulations.

Conventional DFT-based molecular dynamics (DFT-MD) rarely accesses the system sizes and timescales required to capture lithium migration events. Machine learning potentials (MLPs) offer a promising alternative by accelerating computations while retaining near-DFT accuracy\cite{guoAcceleratedAtomisticModeling2021}.

Recently, considerable effort has been devoted to developing MLP frameworks, including Behler–Parrinello neural network,\cite{behlerGeneralizedNeuralNetworkRepresentation2007} the smooth overlap of atomic positions (SOAP) descriptor,\cite{bartokRepresentingChemicalEnvironments2013} the atomistic neural network framework ænet,\cite{artrith2016implementation} Deep Potential Molecular Dynamics (DEEPMD),\cite{zhangDeepPotentialMolecular2018} and MACE.\cite{batatiaDesignSpaceE3Equivariant2022,batatiaMACEHigherOrder2022} Training robust MLPs typically requires large and diverse datasets, which is computationally demanding when based on DFT.

In this work, we present a data-efficient workflow for constructing an interatomic MLP tailored to NMC811. The workflow consists of two main stages: (i) collecting near-ground-state configurations and (ii) employing active learning to efficiently sample non-equilibrium states. Using this approach, we achieved direct MD simulations of lithium self-diffusion in NMC811 with significantly fewer training data than conventional approaches.

\section*{Methods}

\subsection*{Initial DFT Dataset Generation}
All DFT calculations were performed using the Vienna Ab Initio Simulation Package (VASP 6.4.1).\cite{kresseEfficiencyAbinitioTotal1996b,kresseEfficientIterativeSchemes1996} The Perdew–Burke–Ernzerhof (PBE) exchange–correlation functional\cite{perdewGeneralizedGradientApproximation1996} was employed with a plane-wave cutoff of 480~eV and $\Gamma$-point (1~$\times$~1~$\times$~1) sampling. DFT-D3 van der Waals corrections\cite{grimmeConsistentAccurateInitio2010a} were applied in all systems. Strong on-site Coulomb interactions of the transition-metal 3d electrons were treated using the Hubbard-corrected PBE+U functional.\cite{himmetogluHubbardcorrectedDFTEnergy2014} The effective $U_\mathrm{eff}$ values were set to 5.96, 4.84, and 5.14~eV for Ni, Mn, and Co, respectively \cite{zhouFirstprinciplesPredictionRedox2004}. This parameterization has been successfully employed in previous studies on NMC materials.\cite{garciaSurfaceStructureMorphology2017c,schipperStabilizingNickelrichLayered2016}  Spin-polarized calculations were initialized with magnetic moments of 2.0~$\mu_\mathrm{B}$ for Ni, 4.0~$\mu_\mathrm{B}$ for Mn, and 0.0~$\mu_\mathrm{B}$ for Co.

The initial DFT dataset comprised 985 bulk \ce{Li_{1-x}Ni_{0.8}Mn_{0.1}Co_{0.1}O_{2}} structures.

\subsection*{Nudged Elastic Band Calculations}
Nudged elastic band (NEB) calculations were performed using VASP~6.4.1 with the same computational settings as used for generating the DFT training dataset.

NEB calculations were carried out on a \ce{Li_{24}Ni_{48}Mn_{6}Co_{6}O_{120}} supercell with randomly distributed lithium vacancies and transition-metal ions. Five intermediate images were generated via linear interpolation between the initial and final states. The standard NEB method was employed with a spring constant of 5~eV~\AA$^{-2}$. Geometry optimizations were performed with convergence thresholds of $10^{-4}$~eV for total energy and 0.02~eV~\AA$^{-1}$ for atomic forces. The Brillouin zone was sampled using a $\Gamma$-point mesh.

The resulting minimum energy paths were used to benchmark the MLP against DFT in capturing transition-state energetics. A comparison of relative energies along the NEB path is shown in Figure~\ref{fig:5}.

\subsection*{Machine Learning Potential Fine-Tuning}
To refine the interatomic potential for NMC cathode systems, we fine-tuned a small MACE foundation model (\texttt{mace\_agnesi\_small.model}) using a curated dataset of 985 DFT-labeled atomic configurations. Only atomic energies and forces were used as training targets; stress data were excluded. The loss function employed weights of 10.0 for energies and 1.0 for forces to emphasize energy accuracy during fine-tuning. The dataset was split into training and validation subsets with a validation fraction of 0.10. Model training was performed for up to 500 epochs (\texttt{max\_num\_epochs}~=~500) with a batch size of 10. Stochastic weight averaging (SWA) was enabled to improve generalization and stabilize convergence. Training was carried out on an NVIDIA A40 GPU.

\subsection*{Evolutionary Search with \texorpdfstring{\ae}{ae}vo-MACE}
The fine-tuned MACE foundation model was employed as the energy predictor within the \ae vo-MACE software package.\cite{artrithConstructingFirstprinciplesPhase2018a} To explore stable configurations of NMC811, we conducted an atomistic evolutionary search on \ce{Li_{0-60}Ni_{48}Mn_{6}Co_{6}O_{120}}, with atomic sites grouped into three sublattices: (A) oxygen atoms (fixed), (B) lithium layer (variable occupancy), and (C) transition-metal layer (48~Ni, 6~Mn, and 6~Co), corresponding to the nominal stoichiometry of NMC811. All candidate configurations preserved the overall composition and symmetry constraints of the structure.

The evolutionary algorithm used a population size of 15, elitist retention of the top 4 candidates per generation, and a mutation rate of 0.1 to introduce configurational diversity. No predefined structures were included in the initial population. During the search, the total energy of each candidate was evaluated using the fine-tuned MACE potential described above. The use of a machine-learned potential enabled rapid screening of configurational space at near-DFT accuracy while maintaining tractable computational cost.

To construct the convex hull and identify thermodynamically stable phases, the formation energy for each configuration was calculated as:
\begin{equation}
\Delta_f E = E_{\mathrm{Li}_{1-x}\mathrm{MO}_2} - (1-x) E_{\mathrm{LiMO}_2} - x E_{\mathrm{MO}_2}
\end{equation}
where $E_{\mathrm{Li}_{1-x}\mathrm{MO}_2}$ is the total energy of the partially delithiated structure, and $E_{\mathrm{LiMO}_2}$ and $E_{\mathrm{MO}_2}$ are the energies of the fully lithiated and fully delithiated end members, respectively, as shown in Figure S2.

\subsection*{Active Learning}
To efficiently sample structures from MD trajectories, we employed a supervised committee (model ensemble) strategy,\cite{zhangDPGENConcurrentLearning2020a,kongOvercomingSizeLimit2023} in which the near-ground-state dataset served as the starting pool for active learning. Four \ae net models were trained using different random seeds to form the committee, and one of these models was then used to perform MD simulations via the \ae net–LAMMPS interface.\cite{artrithConstructingFirstprinciplesPhase2018a}

All \ae net–LAMMPS simulations were run for 5~ns with a time step of 2~fs, while the temperature was linearly increased from 1000~K to 2000~K to sample configurations with varying degrees of distortion. In each round of active learning, the exploration set, as an initial structure in each \ae net–LAMMPS simulation, included NMC structures spanning SOC values from 0.0 to 1.0 in increments of 0.1. For each SOC, 20 structures were sampled from the MD trajectories, yielding 200 structures per active learning round.

Candidate structures for DFT single-point calculations were selected based on the maximal standard deviation of atomic forces predicted by the ensemble of models. Specifically, for each configuration $\mathcal{R}_t$, the uncertainty $\epsilon_t$ was defined as
\begin{equation}
\epsilon_t = \max_i \sqrt{\left\langle \left\| F_{w,i}(\mathcal{R}_t) - \left\langle F_{w,i}(\mathcal{R}_t) \right\rangle \right\|^2 \right\rangle},
\label{eq:uncertainty}
\end{equation}
where $F_{w,i}(\mathcal{R}_t) = -\nabla_i E_w(\mathcal{R}_t)$ denotes the force on atom $i$ predicted by model $E_w$, and the average $\langle F_{w,i}(\mathcal{R}_t) \rangle$ is computed as the mean prediction over all models in the ensemble,
\begin{equation}
\left\langle F_{w,i}(\mathcal{R}_t) \right\rangle = \frac{1}{N_m} \sum_{\alpha=1}^{N_m} F_{w_\alpha,i}(\mathcal{R}_t).
\label{eq:force-mean}
\end{equation}

In the initial 20 iterations, structures were selected if their force-uncertainty metric satisfied
\begin{equation}
0.05~\mathrm{eV~\AA^{-1}} \leq \epsilon_t < 0.50~\mathrm{eV~\AA^{-1}},
\label{eq:uncertainty-threshold-1}
\end{equation}
and after the 20th iteration, the upper bound was tightened to
\begin{equation}
0.05~\mathrm{eV~\AA^{-1}} \leq \epsilon_t < 0.15~\mathrm{eV~\AA^{-1}},
\label{eq:uncertainty-threshold-2}
\end{equation}
to reduce the inclusion of unphysical structures and improve the quality of the training dataset \cite{zhangDPGENConcurrentLearning2020}.

\subsection*{\ae net–LAMMPS MD Simulations}
Molecular dynamics simulations were performed using the Large-scale Atomic/Molecular Massively Parallel Simulator (LAMMPS). The \ae net potential\cite{artrith2016implementation} was used to describe interatomic interactions, with parameters assigned to Co, Li, Mn, Ni, and O. All simulations were conducted in the canonical (NVT) ensemble using the canonical sampling through velocity rescaling (CSVR) thermostat to maintain a constant temperature. Periodic boundary conditions were applied in all three directions. The simulations used a time step of 2.0~fs.

For mean square displacement (MSD) analysis, large-scale MD simulations were carried out using the final \ae net model obtained after 35 rounds of active learning. Simulations were initialized with five randomly generated \ce{Li_{0–480}Ni_{384}Mn_{48}Co_{48}O_{960}} structures, each adopting the trigonal $R\bar{3}m$ crystal symmetry. Lithium MSD sampling began after a 200~ps equilibration period, and production simulations were performed for 5~ns in the NVT ensemble.

The lithium diffusion coefficient $D$ was extracted using the Einstein relation,
\begin{equation}
D = \frac{1}{6} \lim_{t \to \infty} \frac{\mathrm{d}}{\mathrm{d}t} \left\langle (\delta r)^2 \right\rangle,
\label{eq:einstein}
\end{equation}
where $\left\langle (\delta r)^2 \right\rangle$ denotes the ensemble-averaged MSD of lithium ions. The MSD values were obtained from LAMMPS via the \texttt{compute msd} command applied to lithium atoms, and $D$ was determined by performing a linear fit to the long-time regime of the MSD curve.

\section*{Results and Discussion}

\begin{center}
\includegraphics[width=1.06\textwidth]{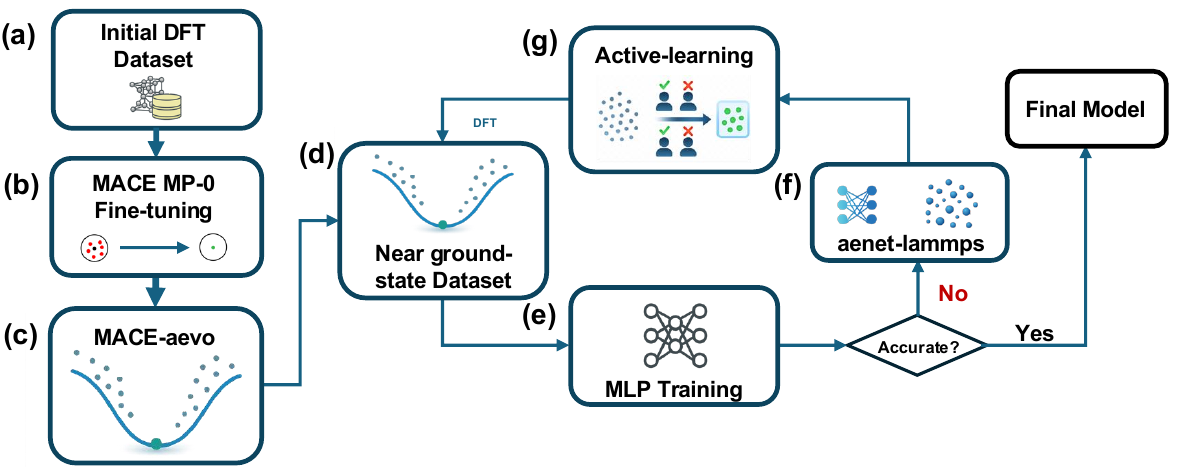}
\captionof{figure}{Overview of the machine learning potential construction workflow. (a) Random \ce{Li_{0-60}Ni_{48}Mn_{6}Co_{6}O_{120}} configurations as the initial dataset; (b) fine-tuned MACE foundation model using the initial DFT dataset; (c) evolutionary search for near-ground-state \ce{Li_{0-60}Ni_{48}Mn_{6}Co_{6}O_{120}} structures using \ae vo-MACE; (d) expanded DFT dataset combining the initial dataset and near-ground-state structures; (e) \ae net MLP training using the dataset from panel~(d); (f) MD simulations using the \ae net potential; (g) active learning to select candidate structures from \ae net–LAMMPS MD trajectories for DFT relabeling.}
\label{fig:1}
\end{center}

Figure~\ref{fig:1} provides an overview of the workflow for constructing an MLP for NMC811. The process can be broadly divided into two stages: the first explores near-ground-state configurations of NMC811, and the second employs active learning to systematically sample non-equilibrium structures. The genetic (evolutionary) algorithm is widely used for sampling near-ground-state structures and has demonstrated strong performance across diverse crystalline materials. In conventional implementations, DFT is typically employed as the energy model, which renders the approach prohibitively expensive for disordered materials with large configurational spaces.

To address this limitation, we employed the pretrained MACE foundation model as a baseline and fine-tuned it on NMC-specific configurations. As shown in Figure~\ref{fig:2}, the original MACE mp-0 model exhibits systematic deviations in both energy and force predictions for a test set of randomly generated \ce{Li_{0–60}Ni_{48}Mn_{6}Co_{6}O_{120}} structures, in which the number and positions of lithium vacancies, as well as the arrangements of Ni, Mn, and Co atoms, are fully randomized. To correct these systematic errors, an additional set of \ce{Li_{0–60}Ni_{48}Mn_{6}Co_{6}O_{120}} configurations was constructed for fine-tuning, with lithium vacancy concentrations of 0, 6, 12, and 18 (evenly spaced in intervals of 6 atoms) to systematically span the compositional space. Single-point DFT calculations were performed on these structures to generate the fine-tuning dataset.

As illustrated in Figure~\ref{fig:2}, the fine-tuning set (highlighted in pink) significantly improved the predictive accuracy of the MACE mp-0 model for disordered NMC configurations, despite the limited amount of DFT data required. These results demonstrate the strong interpolation capability of the MACE foundation model, enabling effective fine-tuning with a small and sparsely distributed dataset to achieve accurate interpolation across the targeted configurational space.

\begin{center}
\includegraphics[width=1.06\textwidth]{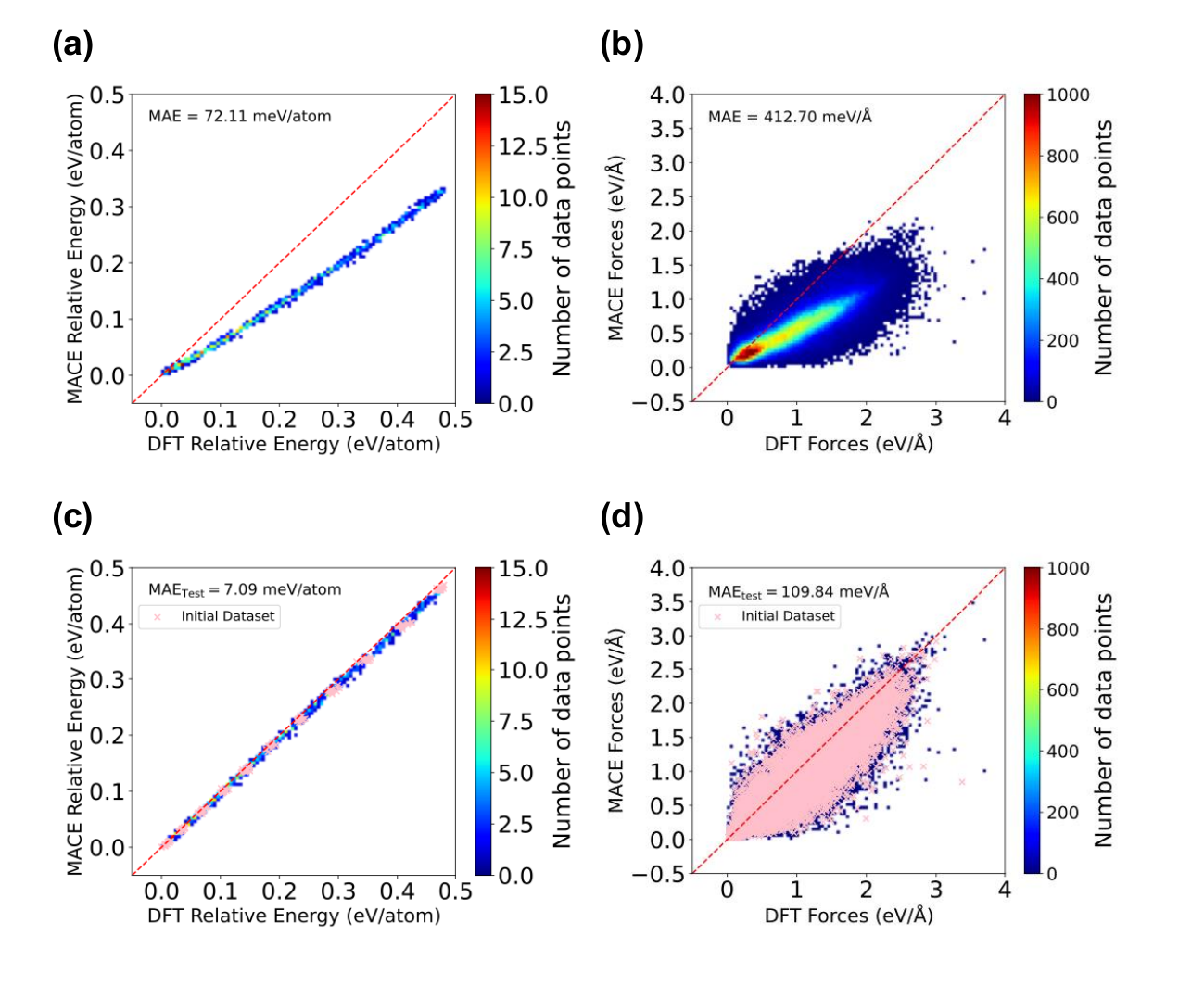}
\captionof{figure}{Fine-tuning of the MACE mp-0 model on randomly generated \ce{Li_{0–60}Ni_{48}Mn_{6}Co_{6}O_{120}} structures. (a,b) Model predictions for energy and atomic forces using the pretrained MACE mp-0 model. (c,d) Model predictions after fine-tuning with 985 DFT-labeled random NMC811 structures.}
\label{fig:2}
\end{center}

After obtaining a fine-tuned model with reliable predictive capability, we applied an evolutionary search to a series of randomly generated \ce{Li_{0–60}Ni_{48}Mn_{6}Co_{6}O_{120}} structures using the \ae vo-MACE package. The resulting near-ground-state structures were further relaxed via DFT geometry optimization. Intermediate configurations from the optimization trajectories were included in the initial dataset, yielding a more diverse NMC811 training set that encompasses a range of structural distortions. However, constructing a robust MLP requires coverage of a broader configurational space, including non-equilibrium structures.

A commonly adopted approach for sampling non-equilibrium structures is to extract configurations from MD trajectories. In previous work, various criteria have been employed to improve the efficiency of sampling by selecting only a subset of configurations for DFT relabeling. For liquid electrolytes, density-matching schemes have been used, while for crystalline materials, force-based selection criteria are commonly applied.\cite{zhangDPGENConcurrentLearning2020a,andradeFreeEnergyProton2020,chenInsightsAtomicMechanism2025,jiaPushingLimitMolecular2020} In the present system, lithium transport behavior depends sensitively on the local potential energy surface, consistent with the octahedral–tetrahedral–octahedral (O–T–O) hopping pathway documented in layered oxides.\cite{jaberiStudyLithiumTransport2024,vandervenFirstprinciplesTheoryIonic2001} Therefore, atomic-force dispersion serves as a suitable criterion for sampling non-equilibrium structures, as it effectively captures the extent of deviation from equilibrium and enables exploration of a broader configurational space.

\begin{center}
\includegraphics[width=1.06\textwidth]{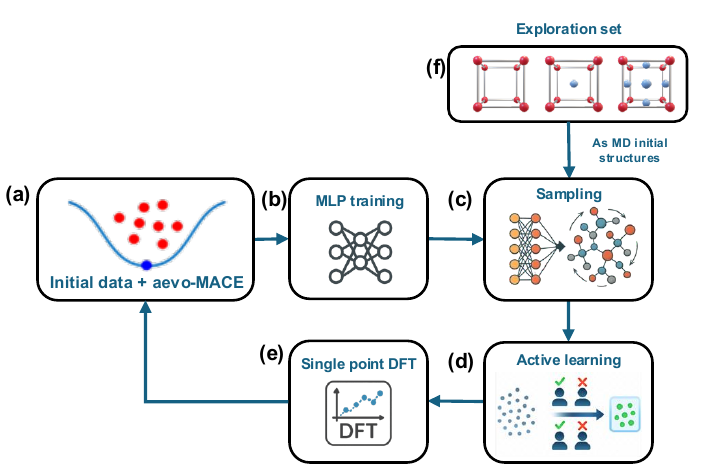}
\captionof{figure}{Schematic illustration of the \ae net-based active learning scheme. (a) DFT dataset from initial random \ce{Li_{0–60}Ni_{48}Mn_{6}Co_{6}O_{120}} structures and near-ground-state structures from \ae vo-MACE; (b) \ae net MLP training; (c) \ae net–LAMMPS MD simulations; (d) active-learning selection of structures from MD trajectories; (e) DFT single-point relabeling; (f) randomly generated initial \ce{Li_{0–60}Ni_{48}Mn_{6}Co_{6}O_{120}} structures used in MD.}
\label{fig:3}
\end{center}

As shown in Figure~\ref{fig:3}a,b, the MLP was trained using both the initial random NMC dataset and the near-ground-state configurations collected via the \ae vo-MACE search. To ensure computational efficiency during active learning, we employed the Chebyshev descriptor within \ae net,\cite{artrithEfficientAccurateMachinelearning2017d} as retraining a full MACE model in each iteration was computationally prohibitive. This MLP was subsequently used to perform \ae net–LAMMPS MD simulations in the 1000–2000~K range to sample non-equilibrium structures. Candidate configurations for DFT relabeling were selected based on the standard deviation of atomic forces predicted by the committee of four active-learning models. Structures with force deviations above the upper threshold were considered unphysical and excluded, whereas those below the lower threshold were deemed already well described by the current MLP. Only structures with intermediate uncertainties were selected for DFT single-point calculations, as shown in Figure~\ref{fig:3}c–e. The candidate structures from MD simulation generally exhibit more diverse energy landscape, as shown in Figure S1. The initial structures for the MD simulations comprised randomly generated NMC configurations with stochastically distributed lithium vacancies and transition-metal arrangements (Figure~\ref{fig:3}f).

\begin{center}
\includegraphics[width=0.75\textwidth]{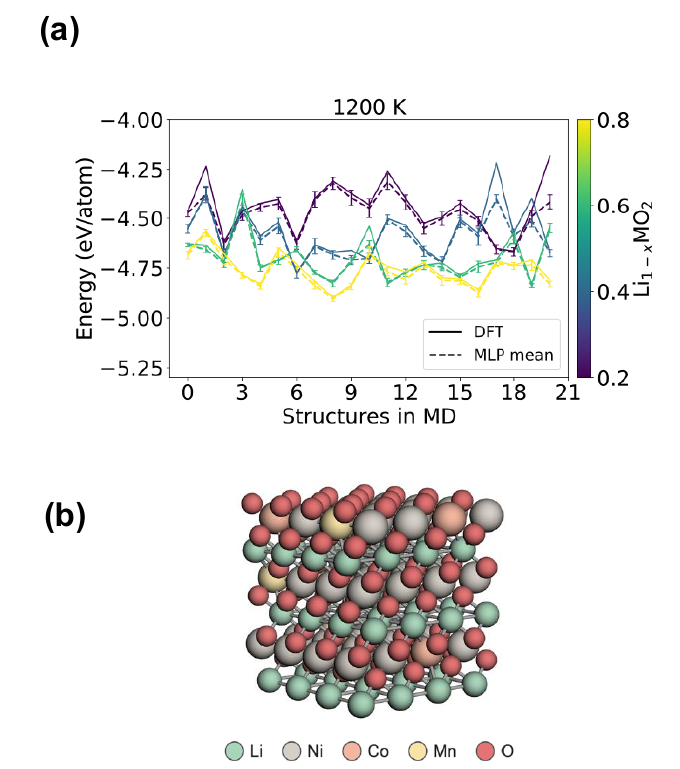}
\captionof{figure}{(a) Energy profiles predicted by DFT (solid lines) and the MLP (dashed lines) for an independent test set of \ce{Li_{0-60}Ni_{48}Mn_{6}Co_{6}O_{120}} configurations along MD trajectories at 1000~K. (b) A representative initial configuration used in the MD. The color bar represents different SOC values.}
\label{fig:4}
\end{center}

To validate the predictive accuracy of the MLP in MD simulations, we evaluated its performance on a set of randomly generated \ce{Li_{0–60}Ni_{48}Mn_{6}Co_{6}O_{120}} structures by comparing the predicted energies along MD trajectories at different temperatures. As shown in Figure~\ref{fig:4}, the MLP after 28 rounds of active learning exhibits excellent predictive performance across temperatures, indicating that the potential has effectively captured a broad region of the underlying potential energy surface.

\begin{center}
\includegraphics[width=1.06\textwidth]{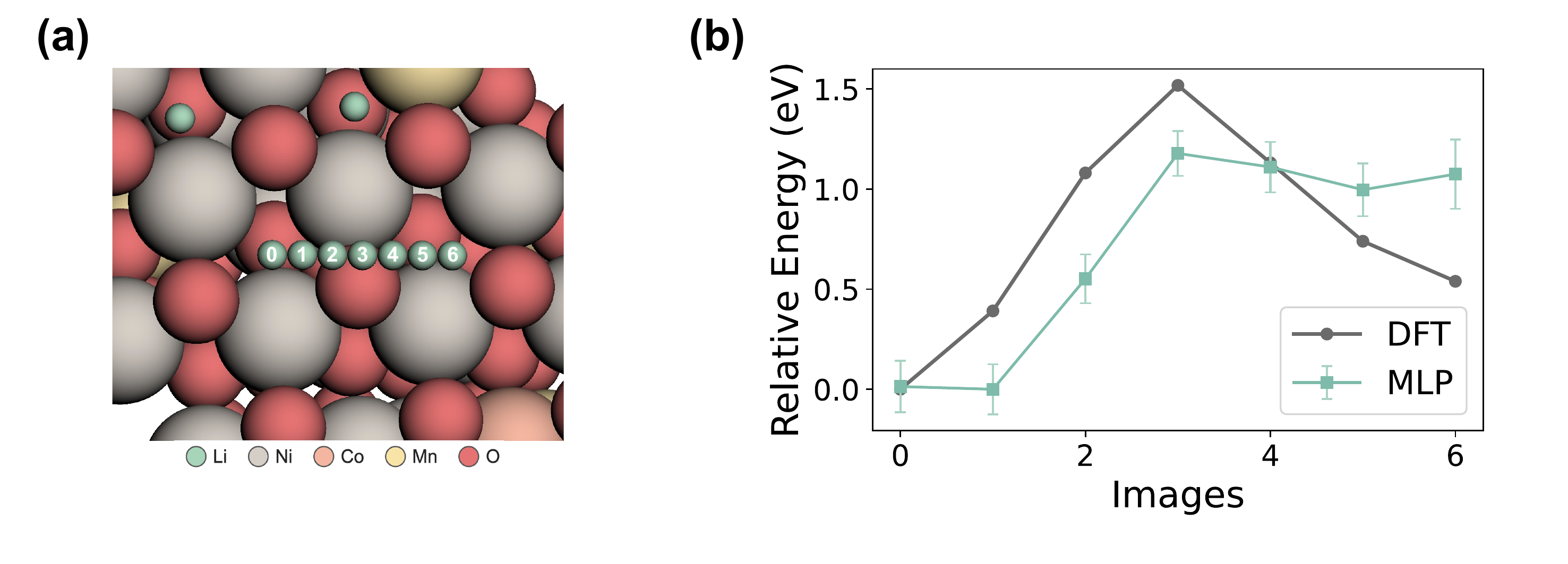}
\captionof{figure}{(a) Schematic illustration of an NEB calculation on a \ce{Li_{24}Ni_{48}Mn_{6}Co_{6}O_{120}} structure with randomly distributed lithium vacancies and transition-metal ions. Images~0 and~6 correspond to the initial and final states of lithium migration, respectively, while images~1–5 represent the NEB path. (b) Comparison of relative energies along the NEB path predicted by DFT and the MLP; the DFT energy of image~0 is set to 0 eV. use average.}
\label{fig:5}
\end{center}

As shown in Figure~\ref{fig:5}a, we performed an NEB calculation on a \ce{Li_{24}Ni_{48}Mn_{6}Co_{6}O_{120}} structure to assess the accuracy of the MLP in describing transition states. Five intermediate images were used to investigate the migration of a lithium ion from an occupied site to a neighboring vacancy. Figure~\ref{fig:5}b presents the energy profiles predicted by the MLP and DFT for the initial, final, and intermediate configurations along the NEB path, demonstrating good agreement between the two methods.

\begin{center}
\includegraphics[width=0.85\textwidth]{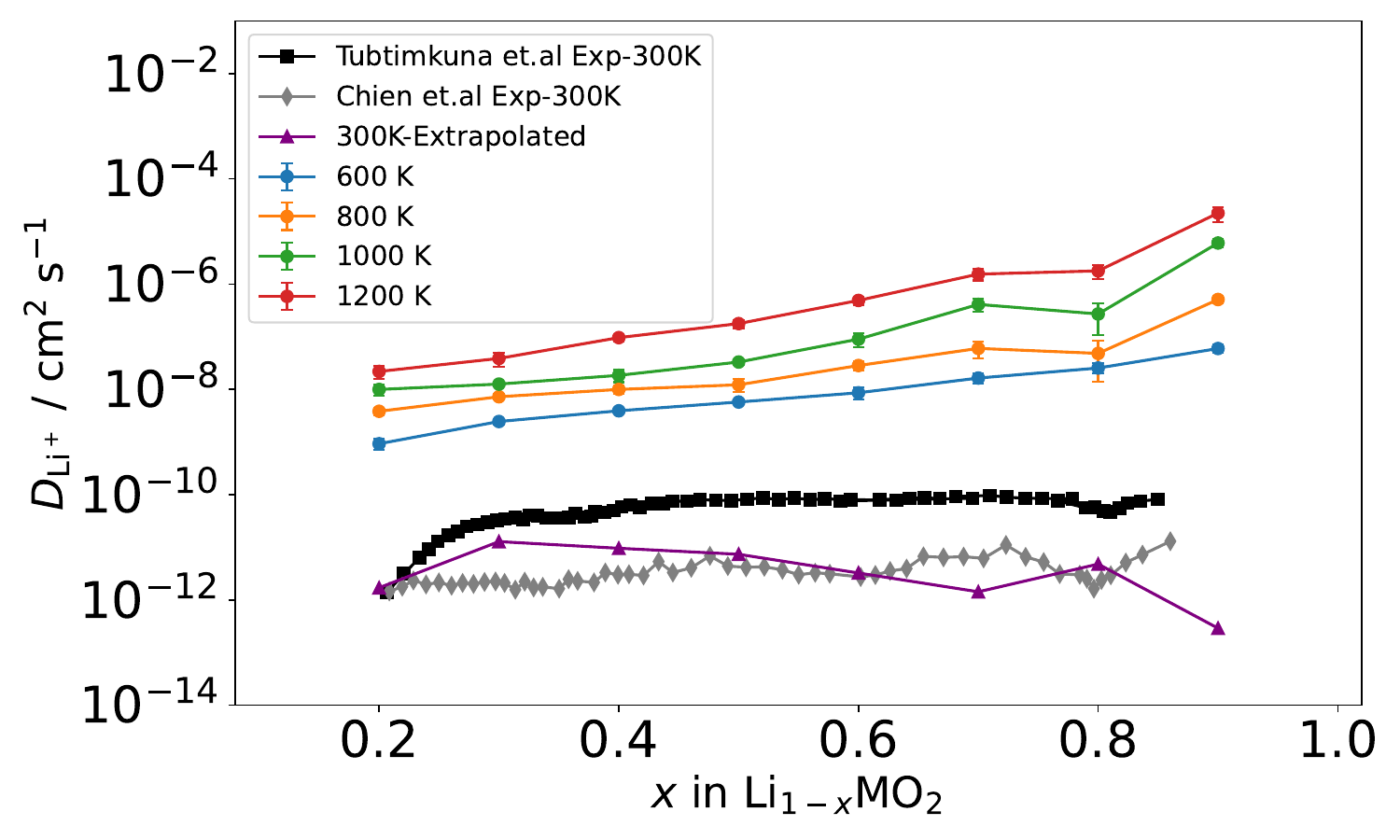}
\captionof{figure}{MLP-predicted lithium self-diffusion coefficients from large-scale \ae net–LAMMPS MD simulations; experimental data are taken from Tubtimkuna \textit{et~al.}\cite{tubtimkunaDiffusionZirconiumIV2022} and Chien \textit{et~al.}\cite{chienRapidDeterminationSolidstate2023} Simulations were repeated five times using different randomly generated \ce{Li_{48–432}Ni_{384}Mn_{48}Co_{48}O_{960}} structures, each with randomized Li-vacancy and transition-metal distributions; error bars indicate the standard deviation across these simulations.} 
\label{fig:6}
\end{center}

\begin{center}
\includegraphics[width=1.0\textwidth]{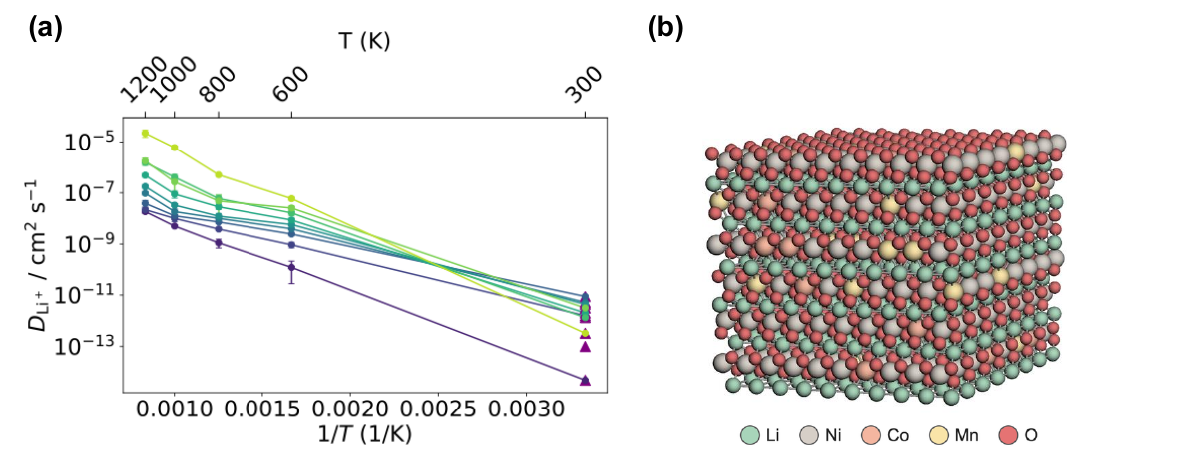}
\captionof{figure}{(a)Diffusion coefficients obtained at elevated temperatures were extrapolated to 300~K using the Arrhenius relation for comparison with experiment. (b) A representative atomic configuration of \ce{Li_{48–432}Ni_{384}Mn_{48}Co_{48}O_{960}} used in the diffusion simulations.} 
\label{fig:7}
\end{center}

After confirming the generalization capability of the MLP, we used it to simulate large-scale lithium self-diffusion in NMC811 with five randomly generated supercells of \ce{Li_{48–432}Ni_{384}Mn_{48}Co_{48}O_{960}}. MD simulations were performed at temperatures ranging from 600 to 1000~K, with a total production time of 5~ns after a 200~ps equilibration period. The MSD was collected for lithium ions, and the diffusion coefficient was calculated via the Einstein relation. Additionally, the Arrhenius relation was used to extrapolate the diffusion coefficient to 300~K for comparison with available experimental values, as shown in Figure~\ref{fig:7} . Detailed MSD data can be found in Figure S4-6.

As shown in Figure~\ref{fig:6}, the MLP-predicted diffusion coefficients exhibit qualitative agreement with experimental data. Notably, at low SOC the predicted diffusion coefficients are closer to experimental values. This trend is likely due to the lower concentration of lithium vacancies and reduced configurational degrees of freedom at low SOC, which yields improved predictive performance under uniform sampling. In contrast, at high SOC the predicted diffusion coefficients show larger variance across the five randomly generated structures, indicating reduced accuracy for systems with higher configurational complexity under uniform sampling conditions.

\section*{Conclusion}
We presented a data-efficient and computationally tractable workflow for constructing a machine learning interatomic potential tailored to the complex, disordered structure of NMC811. By combining a fine-tuned MACE foundation model, an evolutionary algorithm for near-ground-state structure exploration, and an uncertainty-based active learning scheme, we efficiently sampled both equilibrium and non-equilibrium configurations across a wide range of lithium stoichiometries. The resulting potential enabled large-scale MD simulations to directly predict lithium self-diffusion coefficients in NMC811 over a wide temperature range.

Validation against DFT calculations demonstrated the potential’s accuracy in describing both ground-state and transition-state energies. Furthermore, the predicted diffusion coefficients exhibit qualitative agreement with experimental data, especially under high SOC conditions where the configurational complexity is reduced. These results highlight the potential of machine-learning-based approaches to go beyond conventional DFT and cluster-expansion methods by directly simulating transport properties in realistic, disordered cathode materials.

The proposed workflow can be extended to other complex systems and provides a robust framework for understanding ion transport behavior, which is essential for the rational design of high-performance battery materials.

\section*{Data Availability Statement}

The machine learning training datasets and model files used in this work are openly available on the Materials Cloud Archive at \url{https://www.materialscloud.org/xxxx}. Additional data supporting the findings of this study are available from the corresponding author upon reasonable request.

\section*{Conflict of Interest}
The authors declare no competing financial or non-financial interests.

\section*{Acknowledgment}
The authors thank the Dutch National e-Infrastructure and the SURF Cooperative for the computational resources used in the DFT calculations. This work was funded by a start-up grant (Dutch Sector Plan) from Utrecht University awarded to N.A.

% References: ACS style
\bibliography{references_abbrev}

% ============================================================
% Supporting Information
% ============================================================
\clearpage

\setcounter{page}{1}
\renewcommand{\thepage}{S\arabic{page}}
\setcounter{figure}{0}
\setcounter{table}{0}
\setcounter{section}{0}
\renewcommand{\thefigure}{S\arabic{figure}}
\renewcommand{\thetable}{S\arabic{table}}
\renewcommand{\thesection}{S\arabic{section}}

\pagestyle{fancy}
\fancyhf{}
\fancyfoot[C]{\thepage}

\begin{center}
{\Large \textbf{Supporting Information}}\\[1.5em]
{\large Direct Simulation of \ce{LiNi_{0.8}Mn_{0.1}Co_{0.1}O_{2}} Transport Properties Using an Efficient and Accurate Machine Learning Potential}\\[1.5em]
Jian He, Constantijn H. J. A. van de Wetering, Rolande W. Nolsen, Nongnuch Artrith\\[0.5em]
Materials Chemistry and Catalysis, Debye Institute for Nanomaterials Science, Utrecht University, Universiteitsweg 99, 3584 CG Utrecht, The Netherlands\\[0.5em]
\texttt{n.artrith@uu.nl}
\end{center}

\vspace{2em}

\begin{center}
\includegraphics[width=0.7\textwidth]{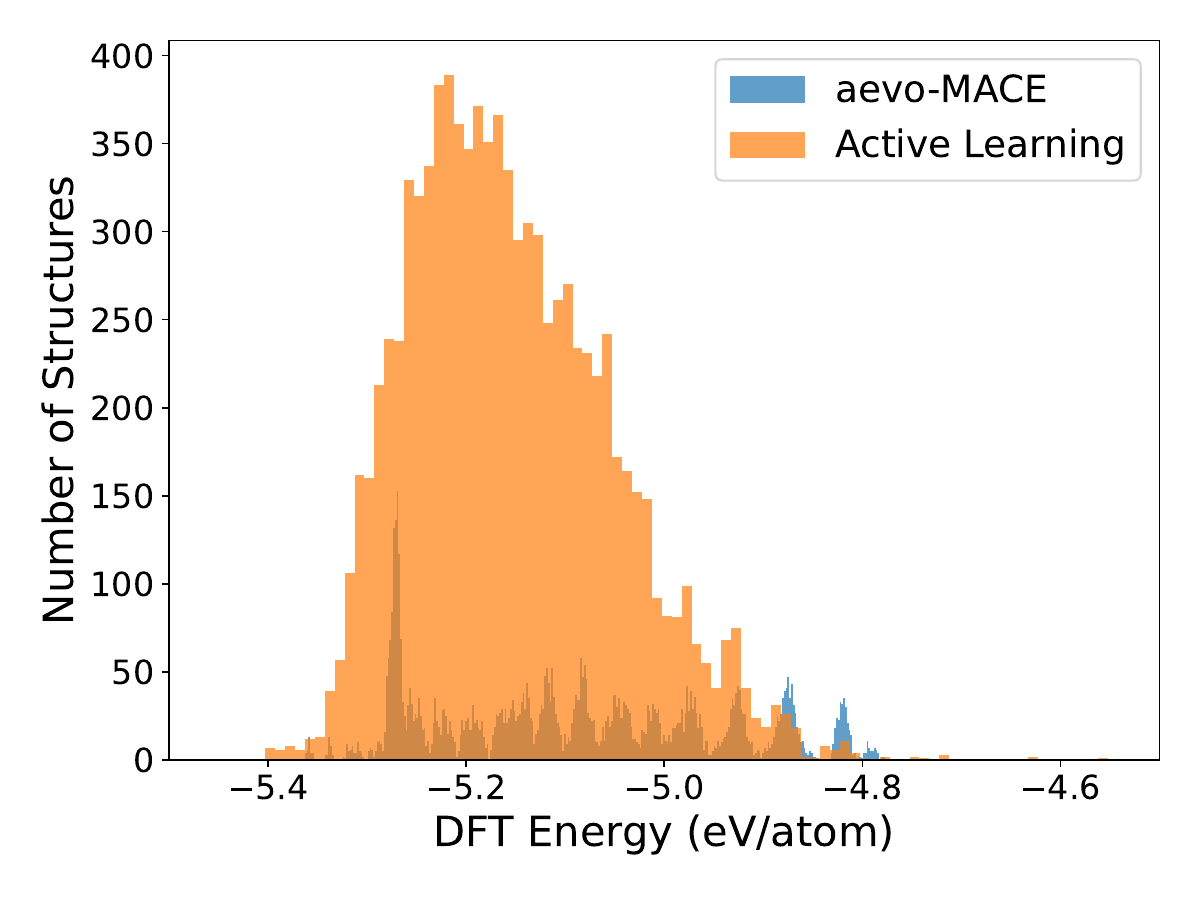}
\captionof{figure}{Energy distribution of training datasets from aevo-MACE and active learning.}
\label{fig:SI1}
\end{center}

\begin{center}
\includegraphics[width=0.7\textwidth]{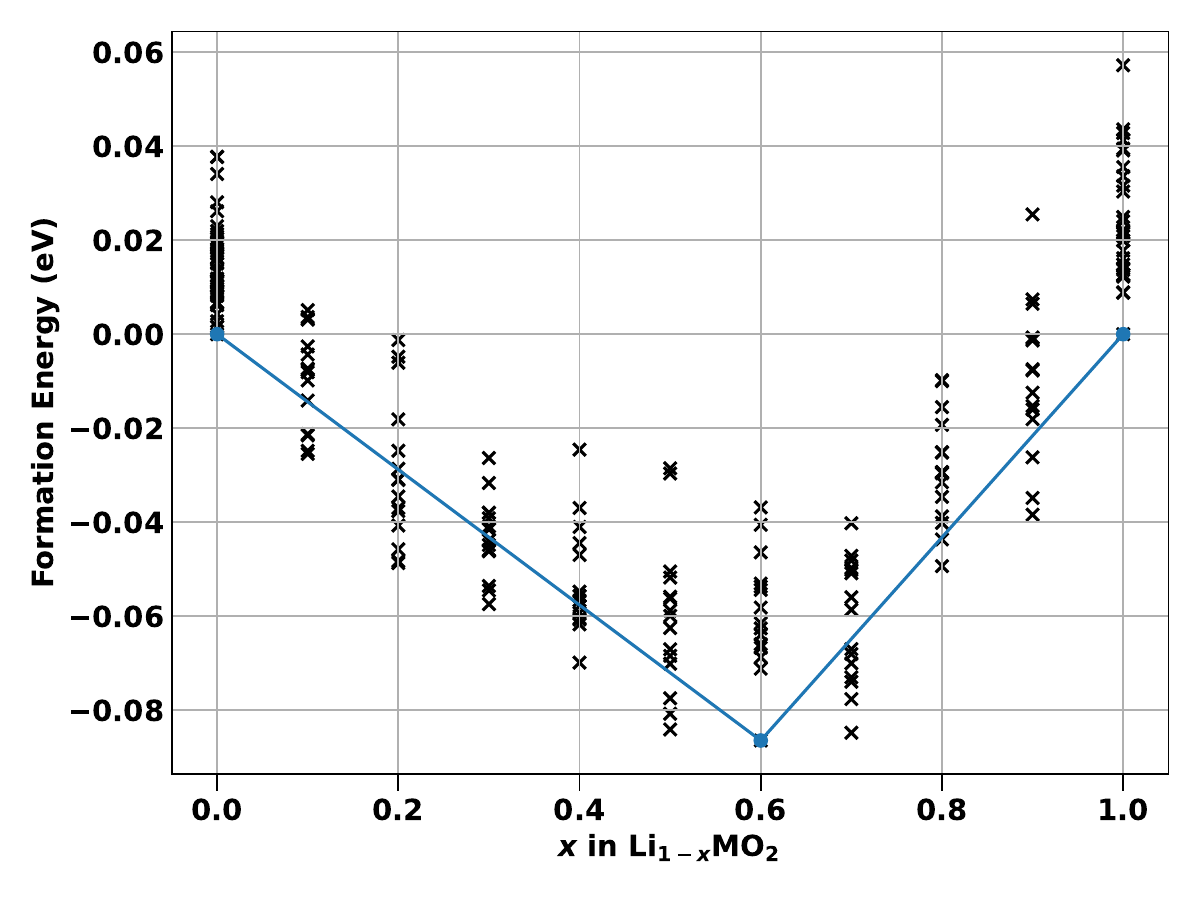}
\captionof{figure}{Convex hull diagram of the \ce{Li_{1-x}Ni_{0.8}Mn_{0.1}Co_{0.1}O_2} system. Atomic configurations were generated using the aevo-MACE genetic algorithm, and formation energies were calculated by DFT. }
\label{fig:SI2}
\end{center}

\begin{center}
\includegraphics[width=0.7\textwidth]{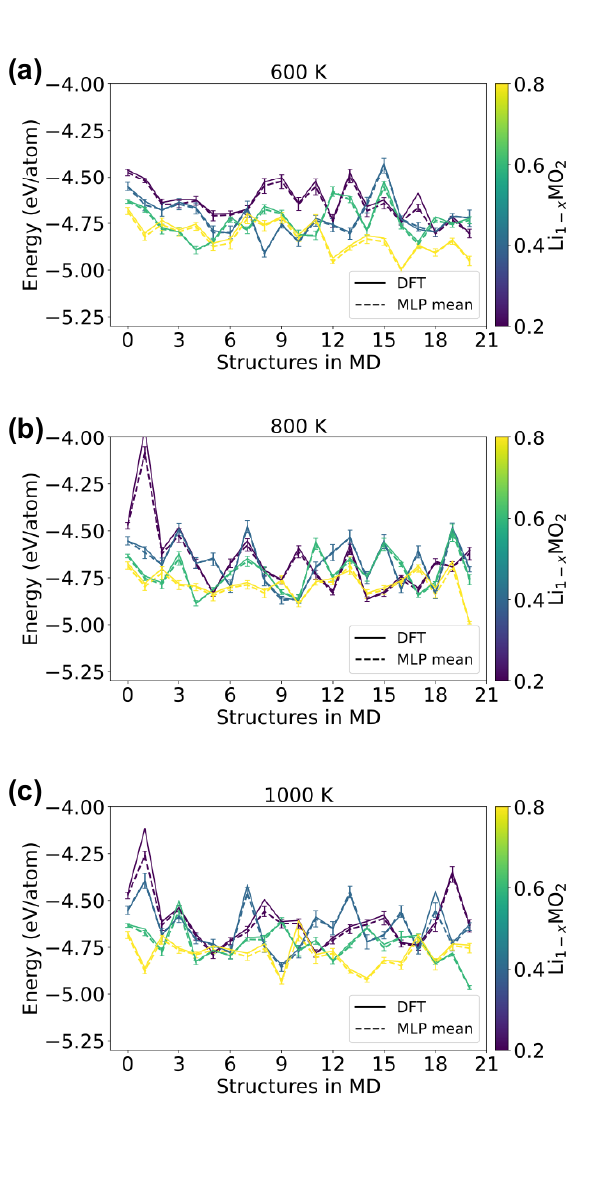}
\captionof{figure}{Energy profiles predicted by DFT (solid lines) and the MLP (dashed lines) for an independent test set of \ce{Li_{0-60}Ni_{48}Mn_{6}Co_{6}O_{120}} configurations along MD trajectories at different temperatures: (a) 600~K, (b) 800~K, (c) 1200~K .}
\label{fig:SI3}
\end{center}

\begin{center}
\includegraphics[width=0.7\textwidth]{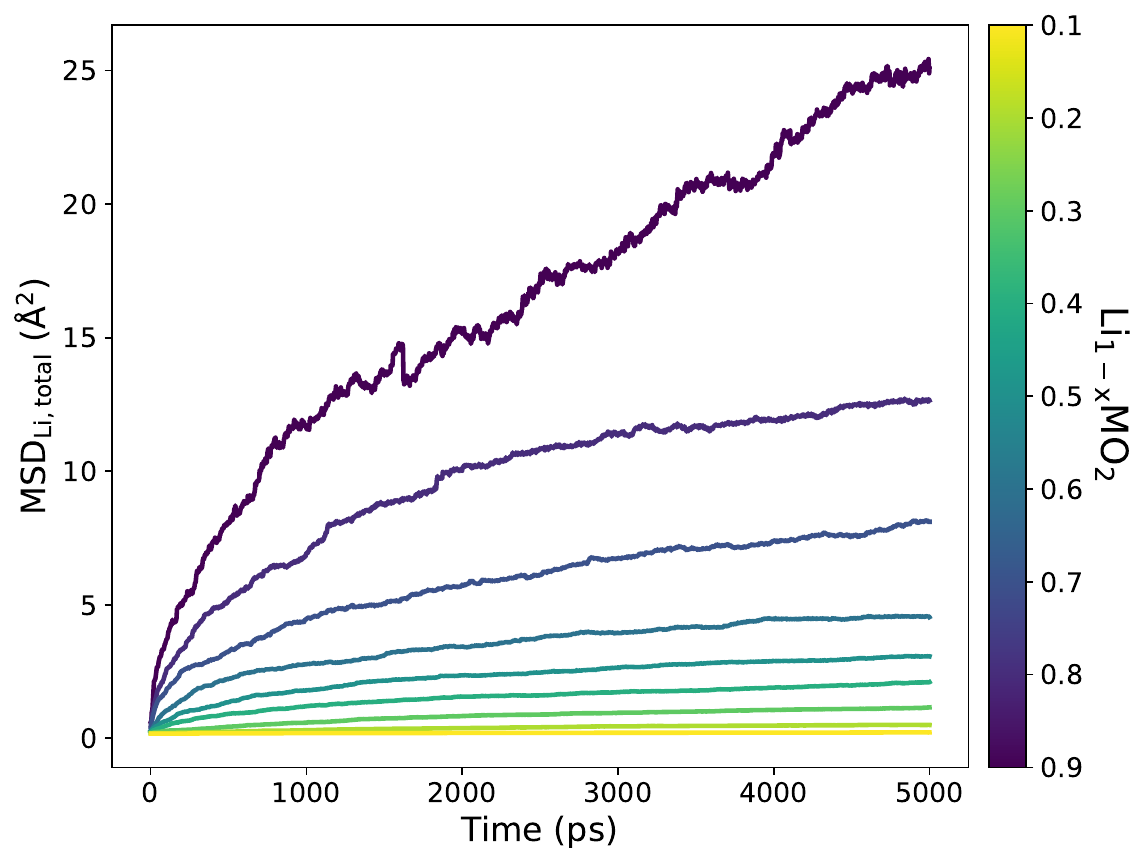}
\captionof{figure}{Li-ion mean square displacement (MSD) profiles at 600 K, averaged over five distinct randomly generated \ce{Li_{48–432}Ni_{384}Mn_{48}Co_{48}O_{960}} configurations in aenet–LAMMPS simulations.}
\label{fig:SI4}
\end{center}

\begin{center}
\includegraphics[width=0.7\textwidth]{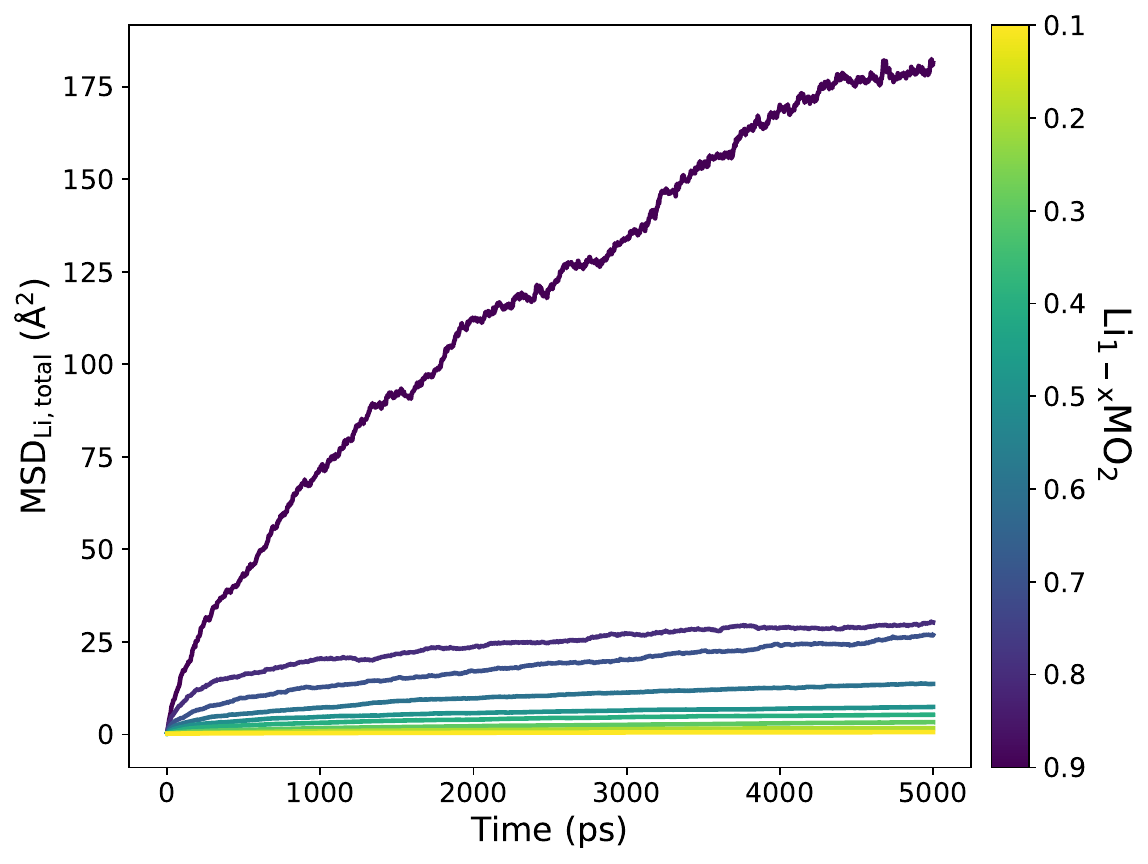}
\captionof{figure}{Li-ion mean square displacement (MSD) profiles at 800 K, averaged over five distinct randomly generated \ce{Li_{48–432}Ni_{384}Mn_{48}Co_{48}O_{960}} configurations in aenet–LAMMPS simulations.}
\label{fig:SI5}
\end{center}

\begin{center}
\includegraphics[width=0.7\textwidth]{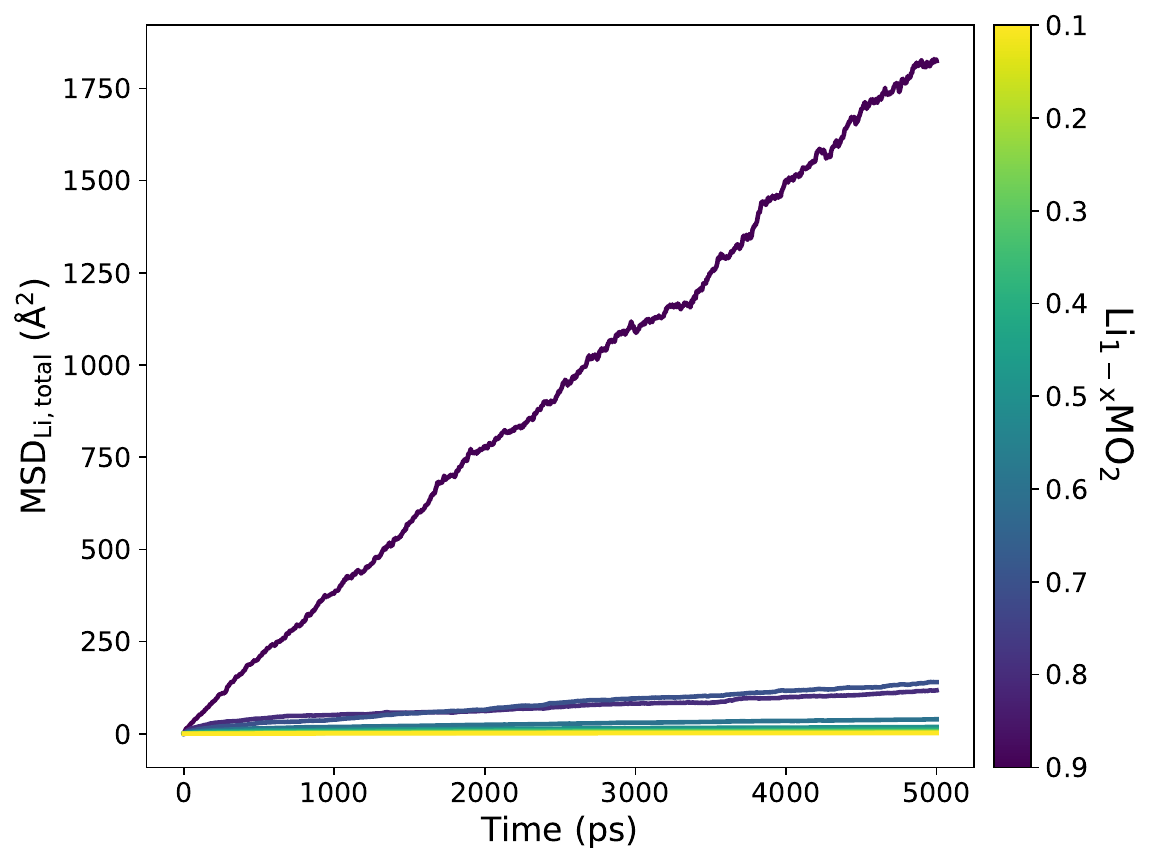}
\captionof{figure}{Li-ion mean square displacement (MSD) profiles at 1000 K, averaged over five distinct randomly generated \ce{Li_{48–432}Ni_{384}Mn_{48}Co_{48}O_{960}} configurations in aenet–LAMMPS simulations.}
\label{fig:SI6}
\end{center}

\begin{center}
\includegraphics[width=0.7\textwidth]{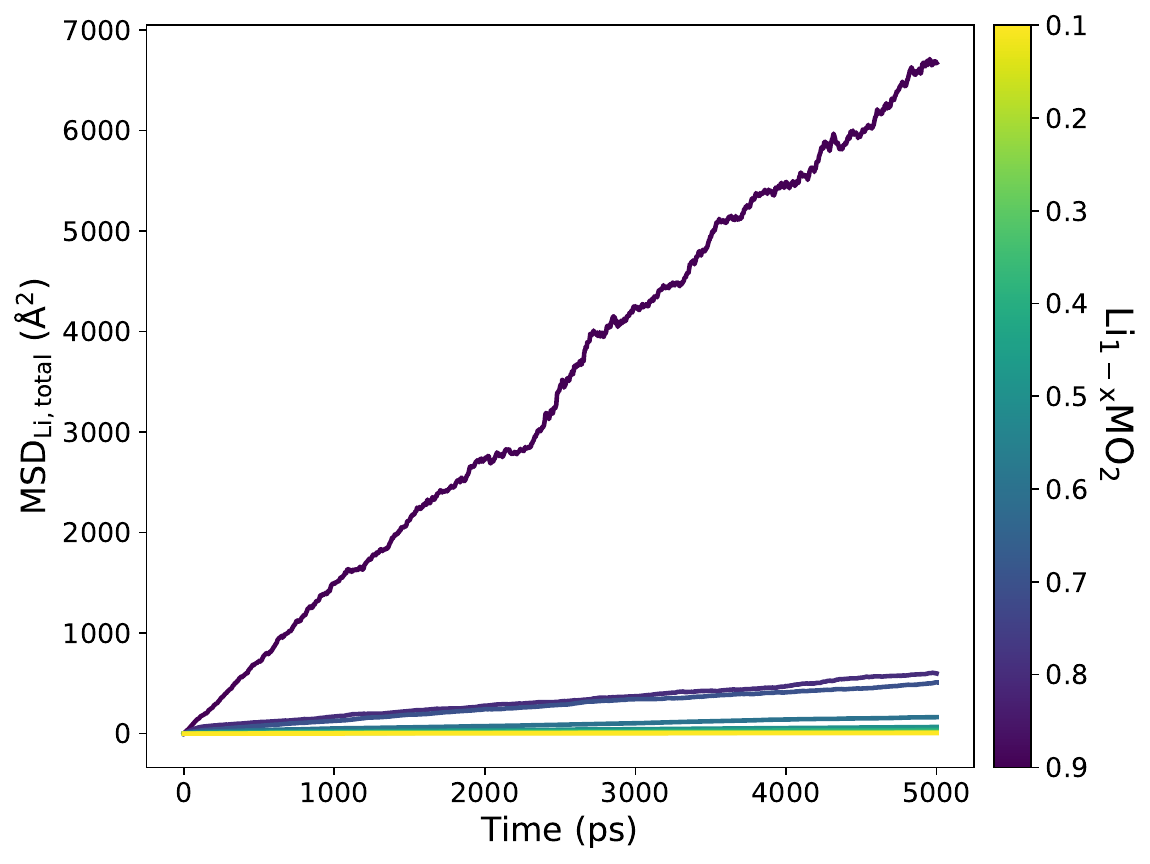}
\captionof{figure}{Li-ion mean square displacement (MSD) profiles at 1200 K, averaged over five distinct randomly generated \ce{Li_{48–432}Ni_{384}Mn_{48}Co_{48}O_{960}} configurations in aenet–LAMMPS simulations.}
\label{fig:SI7}
\end{center}

\end{document}